\documentclass[aip,amsmath,amssymb,reprint]{revtex4-1}

\usepackage{graphicx}
\usepackage{dcolumn}
\usepackage{bm}

\begin{document}

\preprint{}


\title{Extended Magnetic Exchange Interactions in the High-Temperature Ferromagnet MnBi}


\author{T.J.~Williams}
 \email{williamstj@ornl.gov}
 \affiliation{Quantum Condensed Matter Division,
    Neutron Sciences Directorate,
    Oak Ridge National Lab,
    Oak Ridge, TN, 37831, USA}
   
\author{A.E.~Taylor}
 \affiliation{Quantum Condensed Matter Division,
   Neutron Sciences Directorate,
   Oak Ridge National Lab,
   Oak Ridge, TN, 37831, USA}

\author{A.D.~Christianson}
 \affiliation{Quantum Condensed Matter Division,
   Neutron Sciences Directorate,
   Oak Ridge National Lab,
   Oak Ridge, TN, 37831, USA}
 \affiliation{Department of Physics \& Astronomy, 
 	University of Tennessee, 
 	Knoxville, TN, 37966, USA}
 
\author{S.E.~Hahn}
\affiliation{Neutron Data Analysis \& Visualization Division,
	Neutron Sciences Directorate,
	Oak Ridge National Lab,
	Oak Ridge, TN, 37831, USA}

\author{R.S.~Fishman}
\affiliation{Materials Science and Technology Division,
	Physical Sciences Directorate,
	Oak Ridge National Lab,
	Oak Ridge, TN, 37831, USA}

\author{D.S.~Parker}
\affiliation{Materials Science and Technology Division,
	Physical Sciences Directorate,
	Oak Ridge National Lab,
	Oak Ridge, TN, 37831, USA}

\author{M.A.~McGuire}
 \affiliation{Materials Science and Technology Division,
   Physical Sciences Directorate,
   Oak Ridge National Lab,
   Oak Ridge, TN, 37831, USA}
   
\author{B.C.~Sales}
  \affiliation{Materials Science and Technology Division,
   Physical Sciences Directorate,
   Oak Ridge National Lab,
   Oak Ridge, TN, 37831, USA}

\author{M.D.~Lumsden}
 \affiliation{Quantum Condensed Matter Division,
   Neutron Sciences Directorate,
   Oak Ridge National Lab,
   Oak Ridge, TN, 37831, USA}

\date{\today}

\begin{abstract}

The high-temperature ferromagnet MnBi continues to receive attention as a 
candidate to replace rare-earth-containing permanent magnets in applications 
above room temperature.  This is due to a high Curie temperature, large 
magnetic moments, and a coercivity that increases with temperature.  The 
synthesis of MnBi also allows for crystals that are free of interstitial Mn, 
enabling more direct access to the key interactions underlying the physical 
properties of binary Mn-based ferromagnets.  In this work, we use inelastic 
neutron scattering to measure the spin waves of MnBi in order to characterize 
the magnetic exchange at low temperature.  Consistent with the spin 
reorientation that occurs below 140~K, we do not observe a spin gap in this 
system above our experimental resolution.  A Heisenberg model was fit to the 
spin wave data in order to characterize the long-range nature of the 
exchange.  It was found that interactions up to sixth nearest neighbor are 
required to fully parameterize the spin waves.  Surprisingly, the 
nearest-neighbor term is antiferromagnetic, and the realization of a 
ferromagnetic ground state relies on the more numerous ferromagnetic terms 
beyond nearest neighbor, suggesting that the ferromagnetic ground state arises 
as a consequence of the long-ranged interactions in the system. 

\end{abstract}


\maketitle

The unusual magnetic properties of MnBi produce a material that is an 
attractive candidate as an alternative to rare-earth-containing ferromagnets 
in applications at room temperature and above~\cite{Poudyal_13}.  Its high 
transition temperature (T$_C$~=~630~K) and strong magnetic 
anisotropy~\cite{Heusler_04,Roberts_56,Yang_02} combined with large Mn 
moments (3.50(2)~$\mu_B$ at 300~K, and 3.90(2)~$\mu_B$ at 
5~K~\cite{McGuire_14}) produce energy products up to 7.7~MGOe~\cite{Yang_02}.  
Of particular interest for applications, MnBi has a magnetic coercivity that 
increases from $\approx$~1~T at room temperature to $\approx$~2.5~T at 
550~K~\cite{Chen_74,Guo_92,Yang_11}.  It also displays favorable 
magneto-optical properties such as a large Kerr effect, which may be useful as 
a magnetic storage medium in thin films~\cite{Williams_57,Sellmyer_95}.  For 
this reason, there is an ongoing, concerted effort to understand the growth 
conditions to optimize its properties for these applications~\cite{Cui_14}.

Despite the recent interest in MnBi, the microscopic origin of its magnetic 
properties have not been completely determined.   Above 140~K, the 
ordered magnetic moments lie along the $c$-axis, but between 140~K and 90~K 
the spins undergo a continuous rotation away from the $c$-axis so that below 
90~K the spins lie in the $ab$-plane~\cite{Yang_02}.  This transition has been 
identified as a symmetry-lowering structural phase transition driven by 
magnetostriction~\cite{McGuire_14}.  It has been suggested that the Bi atoms 
play an important role in driving this spin-reorientation, as the Bi-Bi 
exchange interactions have been proposed as an origin of the observed 
anisotropic thermal expansion~\cite{McGuire_14}.  Local spin density 
approximation (LSDA) calculations suggest that the spin reorientation occurs 
due to an inversion of the orbital momentum minimum on the Bi 
atom~\cite{Antropov_14}, while a tight-binding 
atomic-sphere-approximation (ASA) model predicts a small, 
antiparallel moment on the Bi atom, which undergoes a simultaneous 
spin reorientation~\cite{Yang_02}.

The loss of magnetic order above 630~K is not a typical magnetic 
transition, but occurs instead due to a decomposition of MnBi into 
Mn$_{1.08}$Bi and elemental Bi~\cite{Chen_73}, both paramagnetic at 
this temperature.  The product Mn$_{1.08}$Bi has excess Mn, where the excess 
Mn atoms occupy 8\% of the interstitial vacancies present in 
MnBi~\cite{Chen_74}.  The relative proximity to the eutectic point (535~K) 
makes the synthesis difficult, which is why most prior work has been performed 
on samples that are Mn$_{1.08}$Bi and/or 
polycrystalline~\cite{Roberts_56,Yang_02,Yang_11}.  Due to coupling of the 
magnetism with the structural decomposition, it has been found that 
application of a 10~T magnetic field can stabilize the MnBi phase up to 
650~K~\cite{Adams_52,Liu_05,Onogi_07}.  By cooling the melt in a magnetic 
field, MnBi crystallites can be obtained, as described 
elsewhere~\cite{McGuire_14,Liu_05,Morikawa_98}.  Other isostructural Mn-based 
ferromagnets, such as MnSb~\cite{Taylor_15}, cannot be synthesized in this 
manner and so they contain some degree of interstitial Mn.  Thus, studying the 
interstitial-free MnBi presents the opportunity to understand the magnetism 
and electronic correlations in the most fundamental form -- coincidentally, 
the member of this family with the highest T$_C$ and largest energy 
product~\cite{Yang_02}.

In this work, we use neutron scattering on a single crystal of MnBi that is 
free of interstitial Mn in order to measure the spin waves at low 
temperature.  These measurements permit us to quantitatively determine the 
exchange interactions in this system, characterizing the degree of extended 
exchange that is present.  Comparing this to other Mn-based binary magnets 
demonstrates the role that the long-ranged interactions play in the magnetic 
properties of this class of materials.

\begin{figure}[thb]
\begin{center}
\includegraphics[angle=0,width=\columnwidth,bb=0 0 699.36 678.28]{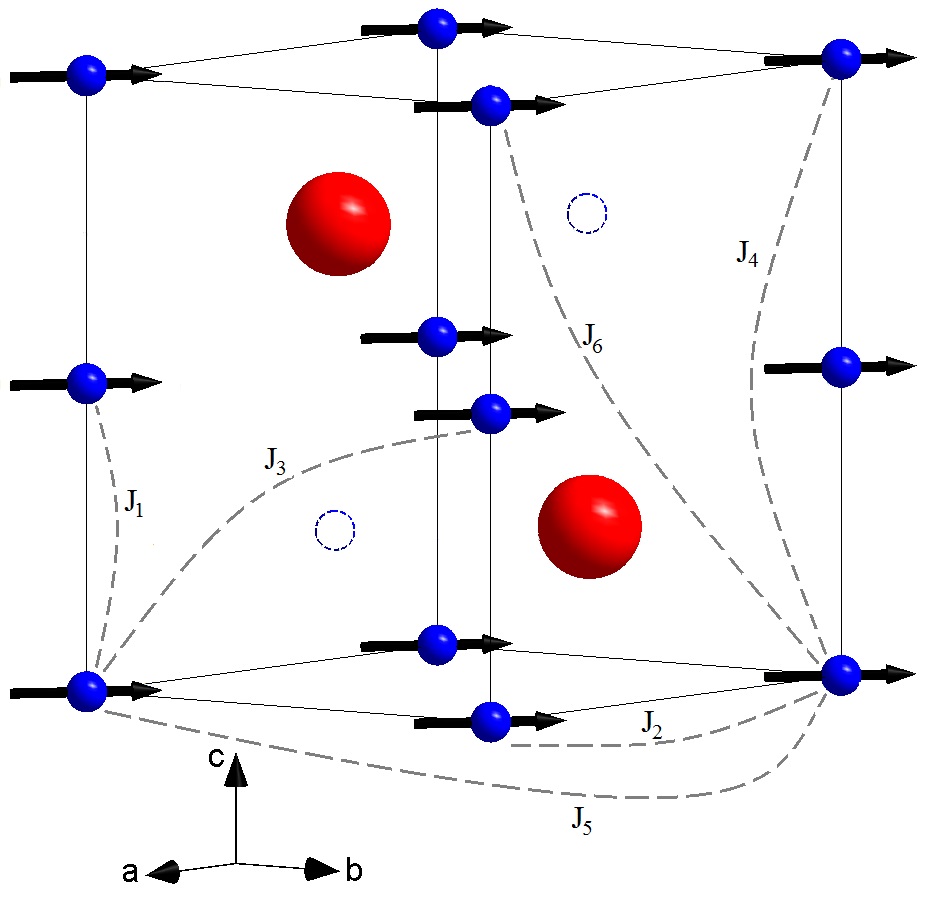}
\caption{\label{structure}
(Color online) The crystal structure of MnBi.  The Mn atoms (small, blue) 
occupy the 2$a$ position, while the Bi atoms (large, red) occupy the 2$c$ 
position.  The other half of the interstitial positions (Wyckoff symbol 2$d$), 
shown as open circles, can be occupied by Mn impurities.  Characterization 
measurements of the samples used suggested that 1(1)~\% of the interstitial 
sites were so occupied by Mn atoms~\cite{McGuire_14}.  The spin arrangement 
shown is for the low-temperature range (T~$<$~90~K).  At all temperatures, 
MnBi is ferromagnetic, however above T~=~140~K, the spins point atlong the 
$c$-axis, while from T~=~140~K to 90~K, the spins rotate away from the 
$c$-axis such that below 90~K the spins lie entirely in the $ab$-plane.  The 
exchange parameters ($J_1$ through $J_6$) determined in this work are shown 
between the relevant Mn ions.
}
\end{center}
\end{figure}

MnBi crystallizes in the hexagonal NiAs-type structure (space group 
P6$_3/mmc$), with the Bi atoms occupying half of the interstitial 
sites~\cite{Roberts_56,Yang_02}.  For this study, single crystals of MnBi were 
grown from a flux utilizing excess Bi, as described 
elsewhere~\cite{McGuire_14}.  The crystals have been characterized 
using powder and single-crystal x-ray diffraction measurements, whose 
refinements gave a nominal composition of Mn$_{1.01(1)}$Bi, where excess 
Mn atoms occupy interstitial sites~\cite{McGuire_14}.  This can be seen in 
Fig.~\ref{structure}, where the open circles denote the interstitial positions 
that can be partially filled by the excess Mn atoms.  

For the neutron scattering measurements, a single crystal of mass 1.989~g and 
a mosaic of 0.41$^{\circ}$ was rotated over a 90$^{\circ}$ range in 
1$^{\circ}$ steps about a vertical axis, with the [$H$ 0 $L$] plane 
horizontal.  The measurements were performed on the Wide Angular-Range Chopper 
Spectrometer (ARCS) at the Spallation Neutron Source (SNS)~\cite{Abernathy_12} 
using incident neutron energies of 80~meV and 150~meV, giving an energy 
resolution of 3.2 and 6~meV, respectively, at the elastic line.  The sample 
was measured in a closed-cycle refrigerator at temperatures of 5~K and 200~K.

\begin{figure*}[thb]
\begin{center}
\includegraphics[angle=0,width=7in,bb=0 0 1542.76 590.21]{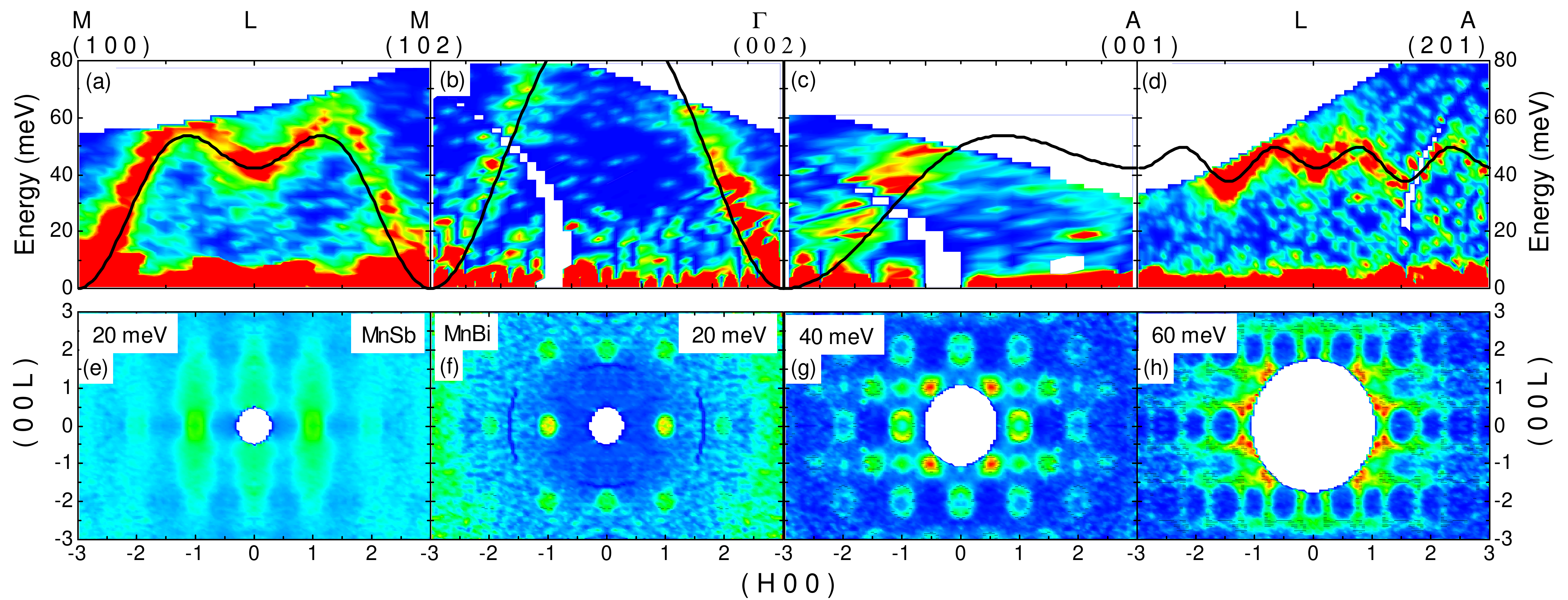}
\caption{\label{sw}
(Color online) Spin wave measurements and fits of MnBi at T~=~5~K.  (a) to 
(d): Inelastic neutron measurements using E$_i$~=~150~meV along different 
reciprocal space directions in the ($H$~0~$L$) scattering plane.  The line is 
a fit to the model described in the text, with the parameters given in 
Table~\ref{exch_tbl}.  Bottom row: Slices of the MnBi data in 
the ($H$~0~$L$) scattering plane at 20~meV (f), 40~meV (g) and 60~meV (h), 
integrating the data $\pm$5~meV.  The inelastic slices show the form factor 
dependence of the spin wave intensity, indicating their magnetic origin.  (e) 
A slice of the related material Mn$_{1.13}$Sb at 20~meV (data taken 
from~\cite{Taylor_15}).  Compared to panel (f), we see that there is 
additional diffuse scattering along $L$ in Mn$_{1.13}$Sb, arising from the 
presence of interstitial Mn.  The lack of similar elastic or inelastic 
signatures is consistent with the MnBi crystal studied being virtually free of 
interstitial Mn.
}
\end{center}
\end{figure*}

The measured magnetic Bragg peaks at 5~K and 200~K are consistent with moments 
directed perpendicular and parallel to the $c$-axis, respectively.  The 5~K 
data, which has reduced phonon scattering due to the Bose factor, was used to 
determine the magnetic exchanges in this system.  The upper panels of 
Fig.~\ref{sw} show representative slices through the spin wave spectrum 
measured on the ARCS spectrometer at T~=~5~K, below the spin reorientation, so 
that the magnetic moments lie within the $ab$-plane.  The spin waves were 
measured across multiple directions in reciprocal space, which allows the 
exchange interactions to be quantified.  We see that the spin waves are 
well-defined, with no broad, inelastic features present in the data that would 
indicate the presence of significant quantities of interstitial Mn, as is 
observed for the case of Mn$_{1.13}$Sb~\cite{Taylor_15} (shown in 
Fig.~\ref{sw}(e)).  This is consistent with the refinement of diffraction data 
that measured 1(1)\% excess Mn in this sample and allows for a simpler 
determination of the exchange interactions since we do not have to account for 
the effects of interstitial Mn on the magnetic exchange.

To parametrize the spin wave dispersion, the data was fit to a Heisenberg 
model.  The moment was assumed to be 3.90~$\mu_B$ on the Mn 
atoms~\cite{McGuire_14}, while no moment was placed on the Bi atoms.  This is 
because the theoretical prediction for the size of the Bi moments is two 
orders of magnitude smaller than the Mn moments, and any moment on the Bi 
sites and accompanying Bi-Bi exchange interactions would be too weak to be 
observed in these measurements~\cite{Yang_02,Antropov_14}.  In order to 
accurately reproduce the spin wave measurements, it was necessary to include 
interactions up to the 6$^{th}$ nearest neighbor ($J_6$), corresponding to a 
distance of 7.462~\AA.  These exchange interactions are shown schematically in 
Fig.~\ref{structure}.  The need to include interactions up to 
$d$~=~7.5~\AA~indicates the significance of relatively long-ranged 
interactions, likely due to an itinerant nature of the magnetic exchange.  
Similarly, other isostructural Mn binary systems MnP~\cite{Obara_80}, 
MnAs~\cite{Menyuk_69, Gama_04}, and MnSb~\cite{Taylor_15,Guillard_49} all have 
large Mn moments and all order ferromagnetically at, or near, ambient 
pressure.  Despite the symmetry-lowering transition at T$_{SR}$~=~90~K, the 
high-temperature space group (NiAs-type) was used for the spin-wave 
calculations as previous measurements of the symmetry-lowering distortion 
indicated that the magnitude of the distortion is very small, and thus 
unlikely to significantly influence the magnetic excitation 
spectrum~\cite{McGuire_14}.  For more details of the fitting procedure, see 
the Supplemental Information~\cite{supp}.

The magnetic anisotropy present in this system necessitates a non-zero spin 
gap.  However, the possible origins of this anisotropy all result in a small 
gap~\cite{McGuire_14,Antropov_14}, which agrees with the measurements of 
Fig~\ref{sw}, where any spin gap is less than the resolution of the data, 
approximately 6~meV.  Measurements on MnSb at low temperatures likewise 
observed no gap, with a resolution of 0.5~meV~\cite{Taylor_15}.  Moreover, 
the system undergoes a spin reorientation between 140~K and 90~K 
suggesting that the magnetic anisotropy is small and strongly 
temperature-dependent~\cite{McGuire_14}.  For these reasons, no spin gap term 
was included when fitting the data.  The line through the data of 
Fig.~\ref{sw} is the fit, showing that the dispersion was able to be 
accurately reproduced with the model that only included the Mn-Mn exchange 
terms given below, without any terms to account for Bi moments or a spin gap, 
though these may still be present.  The lower panels of Fig.~\ref{sw} show 
constant energy slices from the related material Mn$_{1.13}$Sb at 
20~meV~\cite{Taylor_15}, and from MnBi at 20~meV, 40~meV and 60~meV.  
Comparing both 20~meV slices, we see that there is additional scattering in 
Mn$_{1.13}$Sb along the $L$-direction, arising from the presence of 
interstitial Mn.  The lack of that scattering in the MnBi data is consistent 
with the diffraction measurements that have only a small amount of 
interstitial Mn~\cite{McGuire_14}.

\begin{table}[htb]
\begin{center}
\begin{tabular}{|c|c|l|c|l|}
\hline
 Exchange & No. Neighbors & ~~Vector & Distance & Value (meV)
  \\
\hline
$J_{1}$ & ~2 & ~~$c/2$ & ~3.055 \AA~ & ~~~~4.70(17) \\
$J_{2}$ & ~6 & ~~$a$ & 4.283 \AA & ~~~-0.61(10) \\
$J_{3}$ & 12 & ~~$a+c/2$~~ & 5.261 \AA & ~~~-1.73(3) \\
$J_{4}$ & ~2 & ~~$c$ & 6.110 \AA & ~~~-0.12(18) \\
$J_{5}$ & ~6 & ~~$a+b$ & 7.418 \AA & ~~~-1.29(8) \\
$J_{6}$ & 12 & ~~$a+c$ & 7.462 \AA & ~~~-0.63(3) \\
\hline
\end{tabular}
\end{center}
\caption[]{The values for the exchange constants obtained from fitting the 
data in Fig.~\ref{sw}, where positive values denote antiferromagnetic 
exchange.  In order to model the data, it was necessary to include up to 
6$^{th}$ nearest-neighbor interactions due to the extended nature of the 
exchange.}
\label{exch_tbl}
\end{table}

The values of the exchange interactions are shown in Table~\ref{exch_tbl}.  
The effective nearest-neighbor interaction is strongly antiferromagnetic, 
while the rest of the terms are smaller and ferromagnetic.  However, due to 
the larger number of neighbors at greater distances, the ferromagnetic terms 
more than compensate for the $J_1$ term, leading to a ferromagnetic ground 
state.  The antiferromagnetic $J_1$ term is somewhat unexpected, due to the 
large Mn-Mn distance~\cite{Forrer_52,Bai_83}, but a fit that forced $J_1$ to 
be ferromagnetic did not reproduce the data.  As a further consistency check 
on the nature of the magnetic interactions, we have performed first principles 
calculations of the effective Heisenberg exchange parameters using the 
generalized gradient approximation (GGA)~\cite{Perdew_96} in the planewave 
all-electron density functional theory code WIEN2K~\cite{Wien_01} (see the 
Supplemental Information)~\cite{supp}.  In agreement with the experiment, 
$J_1$ is found to be large and antiferromagnetic, but the ground state is 
ferromagnetic due to moderate ferromagnetic values for $J_3$ and $J_5$.  The 
remaining exchange constants are all small, as was found in the fits to the 
experimental data.  The calculations find that the antiferromagnetic state 
lies some 135~meV per Mn higher than the ferromagnetic ground state, leading 
to a mean-field estimated Curie temperature of 522~K, in reasonable agreement 
with experiment.  While the nearest-neighbor exchange term being 
antiferromagnetic is unexpected, both the calculations and the experimental 
data conclude that the long-ranged nature of the Mn exchange give rise 
to ferromagnetism.  This competition between antiferromagnetic local exchange 
and the ferromagnetic itinerant exchange component may also be responsible 
for the reduction of the moment to 3.90~$\mu_B$ from the nominal moment of $g 
S$~=~4~$\mu_B$~\cite{McGuire_14}.  However, the long-range character in MnBi 
clearly dominates, especially when compared to isostructural and isoelectronic 
Mn binary systems (see Table~\ref{mn_struc_params}).

\begin{table}[htb]
\begin{center}
\begin{tabular}{|l|c|c|c|}
\hline
~Compound~ & $T_{Curie}$ (K) & Moment ($\mu_B$) & NN Mn Dist. (\AA)
\\
\hline
 ~MnBi~\cite{McGuire_14} & 630 & 3.90 & 3.055 \\
 ~MnSb~\cite{Elankumaran_92,Radhakrishna_96} & 587 & 3.55 & 2.895 \\
 ~MnAs~\cite{Yuzuri_60,Gama_04} & 318 & 3.20 & 2.852 \\
 ~MnP~\cite{Obara_80,Continenza_01}$^{\star}$ & 292 & 1.33 & 2.743 \\
\hline
\end{tabular}
\end{center}
\caption[]{Magnetic properties of various Mn$T$ ($T$~=~Bi, Sb, As, P) 
materials.  These compounds are found to be ferromagnetic, where the distance 
between the Mn ions increases with increasing size of the $T$ ion.  This 
suggests that there is a systematic increase in the role of the extended 
exchange, which results in higher moments and transition temperatures for 
larger $T$ atoms in the Mn$T$ series.  This agrees well with the results of 
the inelastic neutron scattering on MnBi. $^{\star}$The compound MnP is 
orthorhombic, a slight distortion of the NiAs-type structure of the other 
compounds listed here.  This leads to a helical magnetic state below 50~K, 
however the magnetic structure between 50~K and T$_C$~=~292~K is the same as 
for the other Mn$T$ compounds.}
\label{mn_struc_params}
\end{table}

Of these materials, MnBi has the highest Curie temperature and largest 
ordered moment.  This is likely due to the larger inter-Mn distance, which 
reduces the nearest-neighbor exchange, and places increased emphasis on the 
higher-order exchange terms.  Therefore we expect similar spin wave 
measurements of these compounds to demonstrate a systematic shift from 
long-ranged to more local interactions.  These measurements are greatly 
complicated by the inclusion of interstitial Mn~\cite{Taylor_15}, and it also 
remains an open question to what degree the interstitial ions influence the 
moment and exchange pathways in these various systems~\cite{Radhakrishna_96}.  
However, the systematic changes through the Mn binary ferromagnets are likely 
controlled by the varying importance of the long-range interactions.  


In conclusion, we have observed well-defined spin waves in MnBi.  In order to 
characterize the spin waves, a Heisenberg model with six exchange constants 
has been fit to the data.  The need to include terms out to 
$d$~=~7.5~\AA~suggests that the Mn-Mn exchange is itinerant, and the data 
taken here shows that any gap in this material is less than 6~meV.  The 
resulting fits to the data have shown that the nearest-neighbor term is 
antiferromagnetic, while all of the higher-order terms are ferromagnetic.  
This quantitative determination of the exchange parameters shows that the 
long-range interactions are the determining factor in the ferromagnetic 
ordering, a trend that can explain the reduction in ordered moment and Curie 
temperature in isostructural and isoelectronic Mn binary systems.    

The strength and sign of the exchange constants in MnBi should form a basis 
for comparing to non-stoichiometric Mn$_{1.08}$Bi, as well as other Mn binary 
ferromagnets: MnAs, MnP and MnSb, particularly in understanding the role of 
interstitial Mn on the magnetic properties.  Finally, the results obtained 
here shed light on the role of extended exchange in determining the essential 
characteristics of MnBi for use in permanent magnet applications.

\

We acknowledge instrument support from D.A.~Abernathy and J.~Niedziela.  This 
research at ORNL's Spallation Neutron Source was sponsored by the Scientific 
User Facilities Division, Office of Basic Energy Sciences, US Department of 
Energy.  T.J.W. acknowledges support from the Wigner Fellowship program at Oak 
Ridge National Laboratory.  M.A.M. acknowledges support from U.S. Department 
of Energy, Office of Energy Efficiency and Renewable Energy, Vehicle 
Technologies Office, Propulsion Materials Program.  B.C.S. and D.S.P. were 
supported by the Critical Materials Institute, an Energy Innovation Hub, 
funded by the U. S. Department of Energy, Office of Energy Efficiency and 
Renewable Energy, Advanced Manufacturing Office.  R.S.F. was supported by the 
Department of Energy, Office of Science, Basic Energy Sciences, Materials 
Sciences and Engineering Division.

\clearpage

\addtolength{\oddsidemargin}{-0.75in}
\addtolength{\evensidemargin}{-0.75in}
\addtolength{\topmargin}{-0.725in}

\newcommand{\addpage}[1] {
    \begin{figure*}
      \includegraphics[width=8.5in,page=#1]{MnBi_supplemental.pdf}
    \end{figure*}
}

\addpage{1}
\addpage{2}

\end{document}